\documentclass[11pt]{article}
\usepackage{amsmath,amssymb,epsfig,bm}
\usepackage{caption}
\usepackage{subcaption}

\begin{document}

\title{\bf Peculiar velocities in the $\Lambda$CDM universe}

\author{Evanthia Patliaka${}^1$ and Christos G. Tsagas${}^{1,2}$\\ ${}^1${\small Section of Astrophysics, Astronomy and Mechanics, Department of Physics}\\ {\small Aristotle University of Thessaloniki, Thessaloniki 54124, Greece}\\ ${}^2${\small Clare Hall, University of Cambridge, Herschel Road, Cambridge CB3 9AL, UK}}

\date{\empty}

\maketitle

\begin{abstract}
We present a unified analysis of the linear evolution of peculiar-velocity perturbations in the distribution of pressureless matter after recombination. Our study is carried out within the framework of full general relativity and encompasses both the earlier Einstein–de Sitter epoch and the subsequent $\Lambda$-dominated phase. Starting from a non-interacting, non-comoving mixture of radiation and dust, we derive the generalized differential equations governing these velocity perturbations. We confirm that linear peculiar velocities ($v$) grow as $v\propto t$ throughout the Einstein–de Sitter era. In contrast, during a later phase of increasing $\Omega_{\Lambda}$-contribution, we find that the growth of the peculiar-velocity field is progressively suppressed, before reversing to decay. Nevertheless, part of the earlier velocity growth persists to the present. Hence, bulk-flow velocities at intermediate redshifts are expected to exceed those predicted by the $\Lambda$CDM model, followed by a decline in their value at lower redshifts. Interestingly, such peculiar-velocity profiles have already been reported in the literature.
\end{abstract}

\section{Introduction}\label{sI}
Large-scale peculiar motions are commonplace in the universe. Over the years, numerous surveys have repeatedly confirmed the presence of bulk flows, that is of large regions of the observable universe moving coherently towards a certain direction in the sky (e.g.~see~\cite{ND}-\cite{WHD} for a representative though incomplete list). These surveys seem to agree on the direction of the bulk motions, but to disagree on their size and speed. In fact, some of the reported bulk flows are considerably deeper and faster than those predicted by the current cosmological model, namely by the $\Lambda$CDM paradigm. More specifically surveys confined within the 100/h~Mpc scale seem to largely comply with the $\Lambda$CDM limits~\cite{ND}-\cite{BHL}, while those going beyond this threshold appear to be at odds with it~\cite{WFH}-\cite{WHD}.\footnote{Recall that $h=H/100$~km/secMpc, where $H$ is the Hubble parameter (measured in km/secMpc).} To these, one might also want to add the so-called ``dark flows'', with sizes close to 1000~Mpc and speeds up to 1000~km/sec (e.g.~see~\cite{Ketal}-\cite{AF}). All these bulk motions are believed to have started around the time of recombination as weak peculiar-velocity perturbations, which were subsequently amplified to the observed bulk peculiar flows by structure formation. The interested reader is referred to~\cite{TPA} for a recent review on cosmological peculiar motions, with more details and additional references therein.

It is likely that the fast and deep bulk flows reported by an increasing number of surveys may indicate a generic problem with the $\Lambda$CDM model. However, it is also likely that the same disagreement may reflect a problem in the theoretical models used to study the evolution and the growth of cosmological peculiar velocities. After all, one should always keep in mind that the $\Lambda$CDM predictions are entirely based on Newtonian gravity. The latter predicts the rather mediocre $v\propto t^{1/3}$ growth-rate for the linear peculiar-velocity field ($v$)~\cite{P}. In the literature one may also find a few of the so-called quasi-Newtonian treatments that arrive at the same result~\cite{M,EvEM}. These studies start relativistically but reduce to Newtonian because they impose strict constraints upon the perturbed spacetime, which severely compromise its relativistic nature. The problematic nature and the limitations of the quasi-Newtonian treatment have been known and noted (e.g.~see pages 150 and 151 in \S~6.8.2 of~\cite{EMM} for related warning comments). However, the extent of problem was not realised at the time because there was no direct comparison between a quasi-Newtonian and a proper relativistic study. This was done in a series of recent articles~\cite{TT}-\cite{PT}, which showed that what separates the Newtonian/quasi-Newtonian studies from their relativistic counterparts is that the former bypass, albeit for different reasons, the gravitational input of the \textit{peculiar flux} (see also~\cite{TPA} for further discussion and references).

Peculiar flows are matter in motion and moving matter means nonzero energy flux. In Einstein's theory, as opposed to Newtonian gravity, fluxes also contribute to the energy-momentum tensor and therefore gravitate~\cite{TT}. The gravitational input of this peculiar flux feeds into the Einstein equations and through them emerges in the relations monitoring the peculiar-velocity field. As a result, the latter no longer obeys the Newtonian/quasi-Newtonian linear growth-rate of $v\propto t^{1/3}$. Instead, on an Einstein-de Sitter background, the relativistic analysis leads to a minimum growth-rate of $v\propto t$ for the linear peculiar velocities~\cite{T,PT}. This in turn leads to residual bulk flows considerably faster than anticipated and can potentially reconcile the $\Lambda$CDM model with the fast and deep bulk flows reported in the surveys of~\cite{WFH}-\cite{WHD} for example.

The present work extends the existing relativistic studies by introducing a two-fluid framework and by moving beyond the Einstein–de Sitter epoch. In particular, we generalise the work of~\cite{T,PT} by considering a mixture of radiation and dust, thereby recovering and confirming their results in a more general setting. We also go beyond the Einstein–de Sitter phase by analysing linear peculiar velocities during a late epoch of gradual $\Lambda$-domination. This enables us to assess the impact of an increasing $\Omega_{\Lambda}$-contribution on the evolution of cosmological peculiar velocities. Our results show that a nonzero $\Lambda$ initially suppresses the growth of the peculiar-velocity field and eventually reverses it to decay. The transition appears to occur around the $\Omega_{\Lambda}=1/3$ mark. Below this threshold, the growth of the $v$-field is suppressed, while above it the field decays. Nevertheless, a portion of the earlier velocity growth persists to the present.

Overall, the relativistic analysis indicates a faster growth-rate for the linear peculiar-velocity field in the Einstein–de Sitter era, followed by its suppression and decay once the $\Omega_{\Lambda}$-input becomes appreciable. On this basis, one would expect to measure residual bulk flows faster and deeper than previously anticipated, such as those reported in~\cite{WFH}–\cite{WHD} for example. Moreover, a decline in the bulk velocity at relatively low redshifts should also be expected. Intriguingly, the peculiar-velocity profiles reported in the surveys of~\cite{Cetal,Wetal} appear to exhibit both of these features. It is therefore plausible that the fast and deep bulk flows identified by several surveys do not indicate a fundamental problem with the $\Lambda$CDM model. Rather, they may reflect the use of the inappropriate gravitational theory in the study of cosmological peculiar motions.

\section{Tilted cosmologies}\label{sTCs}
\begin{figure}[tbp]
\centering
\includegraphics[width=0.6\textwidth]{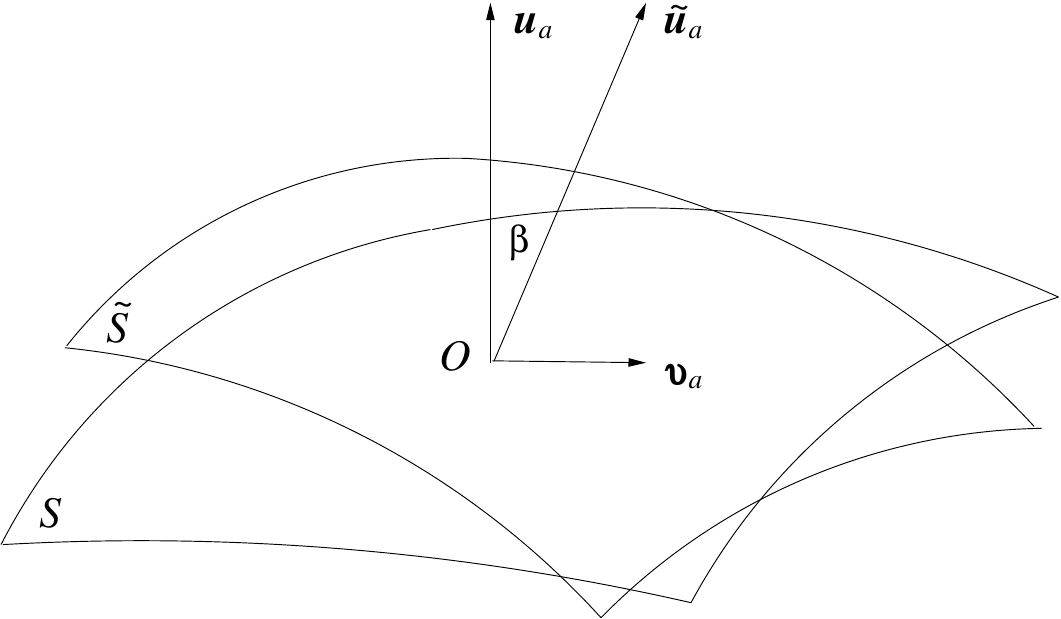}
\caption{Tilted spacetimes with two families of observers/frames in relative motion. The $u_a$-field is identified with the CMB frame, $\tilde{u}_a$ is the 4-velocity of the matter and $v_a$ is its the peculiar velocity. Also, $S$ and $\tilde{S}$ are the 3-D spaces normal to $u_a$ and $\tilde{u}_a$ respectively.}  \label{fig:pmotion}
\end{figure}

Relativistic studies of peculiar motions require ``tilted'' spacetimes, equipped with two (at least) groups of observers in relative motion. The former are the idealised observers following the coordinate system of the Cosmic Microwave Background (CMB), while the latter are identified with the peculiar motion of the matter (e.g.~see Fig.~\ref{fig:pmotion} here and also \S~5 in~\cite{TPA}).

\subsection{Linear relations between the frames}\label{ssLRBFs}
Assuming that the peculiar velocities are not relativistic the three velocity fields seen in Fig.~\ref{fig:pmotion} are related by the familiar (reduced) Lorentz boost $\tilde{u}_a=u_a+v_a$, where $u_au^a=-1=\tilde{u}_a\tilde{u}^a$, $u_av^a=0$, $v^2=v_av^a\ll1$ and $\gamma=(1-v^2)^{1/2}\simeq1$. Then, to linear order, the corresponding projection tensors ($h_{ab}=g_{ab}+u_au_b$ and $\tilde{h}_{ab}=g_{ab}+\tilde{u}_a\tilde{u}_b$ ) satisfy the relation $\tilde{h}_{ab}= h_{ab}+2u_{(a}v_{b)}$. The linear kinematics of the relatively moving observers differ as well. More specifically, on an FRW background, the kinematic variables evaluated in the two frames are related by~\cite{M}
\begin{equation}
\tilde{\Theta}= \Theta+ \vartheta\,, \hspace{5mm} \tilde{\sigma}_{ab}= \sigma_{ab}+ \varsigma_{ab}\,, \hspace{5mm} \tilde{\omega}_{ab}= \omega_{ab}+ \varpi_{ab} \hspace{5mm} {\rm and} \hspace{5mm} \tilde{A}_a= A_a+ \dot{v}_a+ Hv_a\,.  \label{lrels1}
\end{equation}
In the above, $\Theta={\rm D}^au_a$, $\sigma_{ab}={\rm D}_{\langle b}u_{a\rangle}$, $\omega_{ab}={\rm D}_{[b}u_{a]}$ and $A_a=\dot{u}_a$ by definition.\footnote{Overdots indicate time derivatives in the CMB frame, where our analysis will take place.} These are respectively the expansion/contraction scalar, the shear tensor, the vorticity tensor and the 4-acceleration vector measured in the CMB frame, with their tilded counterparts evaluated in the coordinate system of the moving matter. Similarly, $\vartheta={\rm D}^av_a$, $\varsigma_{ab}={\rm D}_{\langle b}v_{a\rangle}$ and $\varpi_{ab}={\rm D}_{[b}v_{a]}$ represent the expansion/contraction, the shear and the vorticity of the peculiar motion.

In relativity, the type of matter varies between observers in relative motion. More specifically, the matter variables in the two coordinate systems satisfy the linear relations~\cite{M}
\begin{equation}
\tilde{\rho}= \rho\,, \hspace{7.5mm} \tilde{p}= p\,, \hspace{7.5mm} \tilde{\pi}_{ab}= \pi_{ab} \hspace{7.5mm} {\rm and} \hspace{7.5mm} \tilde{q}_a= q_a+ (\rho+p)v_a\,.  \label{lrels2}
\end{equation}
Here, $\rho$ is the density, $p$ is the isotropic pressue, $\pi_{ab}$ is the viscosity and $q_a$ is the energy flux of the matter in the CMB frame, with the tilded variables evaluated in the matter frame.

Before closing this section, it is worth drawing the readers attention to expressions (\ref{lrels1}d) and (\ref{lrels2}d) above. The former ensures that, in the presence of peculiar motions, there is nonzero 4-acceleration, irrespective of the presence of pressure. Indeed, even when the 4-acceleration vanishes in one frame, it is nonzero in any other relatively moving coordinate system. Note that setting both $\tilde{A}_a$ and $A_a$ to zero in Eq.~(\ref{lrels1}a) simultaneously, leads to $\dot{v}_a+Hv_a=0$. This in turn leads to peculiar velocities decaying as $v_a\propto 1/a$, since $H=\dot{a}/a$ (with $a$ being the cosmological scale factor). Clearly, given the widespread presence of fast and deep bulk flows in the universe, a decaying peculiar-velocity field is observationally unacceptable.

In an analogous way, expression (\ref{lrels2}d) directly relates the presence of peculiar velocities with that of nonzero energy flux. Put another way, when peculiar velocities are present, the cosmic medium can not be treated as perfect and the imperfection appears in the form of a nonzero ``peculiar flux''~\cite{T}. Indeed, setting both $\tilde{q}_a$ and $q_a$ to zero simultaneously, erases the peculiar-velocity field as well. The only exception occurs when the matter has a de~Sitter inflationary equation of state with $p=-\rho$ (see~\cite{MaT} for an application and further discussion).

\section{Linear evolution of peculiar velocities}\label{sLEPVs}
Velocity perturbations in the baryonic matter are expected to develop after recombination, once the baryons have decoupled from the radiation field and can start agglomerating gravitationally. On the other hand, peculiar velocities in the Cold Dark Matter (CDM) sector could have started developing at an earlier time, depending on the specifics of the still elusive CDM species.

\subsection{A two-fluid analysis}\label{ssTFA}
Our main aim is to study the linear evolution of peculiar-velocity perturbations in the $\Lambda$CDM universe, namely after recombination in the Einstein-de Sitter epoch and then through a  subsequent phase of gradual $\Lambda$-domination. Nevertheless, we will start our analysis by considering a mixture of radiation and dust (baryonic or/and CDM), with densities $\rho^{(r)}$ and $\rho^{(d)}$ respectively. We will also endow the two fluids with (non-relativistic) peculiar velocities $v_a^{(r)}$ and $v_a^{(d)}$ relative to the CMB frame (with 4-velocity $u_a$ -- see Fig.~\ref{fig:pmotion} above). All this will generalise the analysis of~\cite{T,PT} and, in so doing, also provide an alternative derivation of their equations and results.

Let us consider a perturbed Friedmann-Robertson-Walker (FRW) universe containing a mixture of non-interacting radiation and dust. To linear order, peculiar-velocity perturbations in the pressure-free matter is monitored by a system of two differential equations. These monitor linear inhomogeneiies in the matter density ($\Delta^{(d)}_a=(a/\rho^{(d)}){\rm D}_a\rho^{(d)}$) and in the universal expansion ($\mathcal{Z}_a=a{\rm D}_a\Theta$) and they are respectively given by (e.g.~see~\cite{TCM})
\begin{equation}
\dot{\Delta}^{(d)}_a= -\mathcal{Z}_a+ {3aH\over\rho^{(d)}}\left(\dot{q}_a^{(d)}+4Hq_a^{(d)}\right)- {a\over\rho^{(d)}}\,{\rm D}_a{\rm D}^bq^{(d)}_a  \label{dotmDel1}
\end{equation}
and
\begin{equation}
\dot{\mathcal{Z}}_a= -2H\mathcal{Z}_a -{1\over2}\,\rho\Delta_a- {c_s^2\over1+w}\,{\rm D}_a{\rm D}^b\Delta_b+ {3\over2}\,a\left(\dot{q}_a+4Hq_a\right)- {a\over\rho(1+w)}\,{\rm D}_a{\rm D}^b\left(\dot{q}_b+4Hq_b\right)\,.  \label{dotcZ1}
\end{equation}
In the above, $\rho=\rho^{(r)}+\rho^{(d)}$ is the total energy density of the mixture and $q_a=q_a^{(r)}+q_a^{(d)}$ is the total (peculiar) flux, with $q_a^{(r)}=4\rho^{(r)}v_a^{(r)}/3$ being the contribution of the radiative fluid and $q^{(d)}_a=\rho^{(d)}v^{(d)}_a$ that of the dust component. In an analogous way, $\Delta_a=(a/\rho){\rm D}_a\rho$ is the spatial gradient of the total energy density, with
\begin{equation}
\Delta_a= {1\over\rho}\left(\rho^{(r)}\Delta_a^{(r)} +\rho^{(d)}\Delta_a^{(d)}\right)  \label{tDel}
\end{equation}
and $\Delta_a^{(r)}=(a/\rho^{(r)}){\rm D}_a\rho^{(r)}$ by construction. Finally, $w$ and $c_s^2$ are the effective barotropic index and the (squared) sound-speed of the radiation-dust mixture, respectively given by
\begin{equation}
w= {\rho^{(r)}\over3(\rho^{(r)}+\rho^{(d)})} \hspace{15mm} {\rm and} \hspace{15mm} c_s^2= {4\rho^{(r)}\over3(4\rho^{(r)}+3\rho^{(d)})}\,,  \label{twcs}
\end{equation}
keeping in mind that $p^{(r)}=\rho^{(r)}/3$ and $p^{(d)}=0$. As expected the above reduce to $w=1/3=c_s^2$ during the radiation era (when $\rho^{(d)}\ll\rho^{(r)}$), while they approach $w=0=c_s^2$ deep into the dust epoch (when $\rho^{(r)}\ll\rho^{(d)}$). Note that expressions (\ref{dotmDel1}) and (\ref{dotcZ1}) have been obtained after using the momentum conservation equation of the total fluid, namely (e.g.~see~\cite{TCM})
\begin{equation}
a\rho(1+w)A_a= -c_s^2\rho\,\Delta_a- a\left(\dot{q}_a+4Hq_a\right)\,.  \label{tmcl1}
\end{equation}
Therefore, in fact, the relativistic system comprises three rather than two differential equations. Before closing this section, it is worth reminding the reader that the evolution of the expansion gradients depends on the total fluid (see Eq.~(\ref{dotcZ1}) above). This is only natural, since the dynamics of the universal expansion are determined by all the  components of the matter content.

\subsection{The role of the peculiar flux}\label{ssRPF}
The flux-related terms on the right-hand sides of Eqs.~(\ref{dotmDel1}) and (\ref{dotcZ1}) reflect the contribution of the moving matter (radiation and dust) to the relativistic gravitational field. As demonstrated in a series of articles (e.g.~see~\cite{TT}-\cite{PT}, as well as \S~6.3.2 in~\cite{TPA} and \S~\ref{sI} here), the flux-input is what makes the difference between the proper relativistic studies of peculiar motions and the rest.

Bypassing the gravitational contribution of the peculiar flux, for one reason or another, has serious negative side-effects. For example, the quasi-Newtonian studies of~\cite{M,EvEM} reduce to purely Newtonian when the flux effects are not accounted for (see~\cite{T,PT}, as well as \S~6.2 in~\cite{TPA}).\footnote{An additional example is the relativistic study of the Zeldovich approximation~\cite{ET}. The latter simply reproduced the standard purely Newtonian picture, once the quasi-Newtonian approximation was introduced.} In the literature, there are also two-fluid studies that originally incorporate the above flux-terms in their equations, but then switched the analysis to the Landau-Lifshitz frame (e.g.~see \S~10.3.2 in~\cite{EMM}), or to the centre-of-mass frame~\cite{CMV-R}. There, the total energy flux ($q_a$) vanishes by default, leaving Eq.~(\ref{dotcZ1}) without any flux terms whatsoever. The only flux that survived was that of the pressure-free matter ($q_a^{(d)}$) at the end of Eq.~(\ref{dotmDel1}).\footnote{Setting the energy flux to zero does not imply switching the peculiar motion of the matter off. Instead, the energy flux appears now as particle flux. Recall that the Landau-Lifshitz frame has zero energy flux, but nonzero particle flux. This is reversed in the Eckart frame, which has zero particle flux but nonzero energy flux~\cite{TPA}.} Without properly accounting for the flux effects, these studies simply reproduced the purely Newtonian results, despite their initial relativistic setup. In addition, the linear $v^{(d)}$-field was found to decay as $v^{(d)}\propto1/a$. This result is observationally unacceptable, since a decaying $v$-field cannot explain the plethora of surveys reporting fast and deep bulk peculiar flows today.

In what follows, we will assume that the radiation  field is homogeneously distributed and follows the frame of the CMB photons. In other words, we will set $\Delta_a^{(r)}=0=v_a^{(r)}$, in which case $\Delta_a=(\rho^{(d)}/\rho)\Delta_a^{(d)}$ and $q_a=q_a^{(d)}=\rho^{(d)}v_a^{(d)}$. Then, keeping in mind that $\dot{\rho}^{(d)}=-3H\rho^{(d)}$ in the background and using the auxiliary relations (\ref{twcs}), expressions (\ref{dotmDel1}) and (\ref{dotcZ1}) recast into
\begin{equation}
\dot{v}^{(d)}_a+Hv^{(d)}_a= {1\over3aH}\left(\dot{\Delta}^{(d)}_a+\mathcal{Z}_a\right)+ {1\over3H}\,{\rm D}_a\vartheta^{(d)}  \label{dotv}
\end{equation}
and
\begin{eqnarray}
\dot{\mathcal{Z}}_a&=& -2H\mathcal{Z}_a -{1\over2}\,\rho^{(d)}\Delta^{(d)}_a- {4\rho^{(r)}\rho^{(d)}\over(4\rho^{(r)}+4\rho^{(d)})^2}\, {\rm D}_a{\rm D}^b\Delta^{(d)}_b+ {3\over2}\,a\rho^{(d)}\left(\dot{v}^{(d)}_a+Hv^{(d)}_a\right) \nonumber\\ &&-{3a\rho^{(d)}\over4\rho^{(r)}+3\rho^{(d)}}\,{\rm D}_a\dot{\vartheta}^{(d)}- {6aH\rho^{(d)}\over4\rho^{(r)}+3\rho^{(d)}}\,{\rm D}_a\vartheta^{(d)}\,,  \label{dotcZ3}
\end{eqnarray}
having set $\vartheta^{(d)}={\rm D}^av_a^{(d)}$ and used the linear commutation law ${\rm D}^a\dot{v}_a= ({\rm D}^av_a)^{\cdot}+H{\rm D}^av_a$ between spatial gradients and temporal derivatives of first-order spacelike vectors.

Equations (\ref{dotv}) and (\ref{dotcZ3}) are fully relativistic and monitor the linear evolution of peculiar-velocity perturbations in the dust component of an almost-FRW cosmology filled with a non-interacting mixture of radiation and dust. Crucially, both formulae properly incorporate the flux effects triggered by the peculiar motion of the dust. In particular, the last term on the right-hand side of (\ref{dotv}) and the last three terms in (\ref{dotcZ3}) are due to the peculiar flux and carry the latter's contribution to the relativistic gravitational field. Incorporating the flux input is what separates the relativistic study of peculiar motions from the Newtonian/quasi-Newtonian ones. Next, we will apply the above system to the Einstein-de Sitter epoch of the universe.

\section{Peculiar velocities after recombination}\label{sPVAR}
The system (\ref{dotv}), (\ref{dotcZ3}) reveals that the gravitational input of the peculiar flux also brings into play the density and the expansion gradients. Put another way, linear peculiar velocities are driven structure formation and by the increasing inhomogeneity of the post-recombination universe.

\subsection{The Einstein-de Sitter epoch}\label{ssEdSE}
Once deep into the Einstein-de Sitter era, the pressureless matter dominates the energy density of the universe. In practice, this translates into the conditions $\rho^{(d)}\simeq\rho$ and $\rho^{(r)}/\rho^{(d)}\ll1$, in which case Eqs.~(\ref{dotv}) and (\ref{dotcZ3}) reduce to
\begin{equation}
\dot{v}^{(d)}_a+Hv^{(d)}_a= {1\over3aH}\left(\dot{\Delta}^{(d)}_a+\mathcal{Z}_a\right)+ {1\over3H}\,{\rm D}_a\vartheta^{(d)}  \label{EdSdotv}
\end{equation}
and
\begin{equation}
\dot{\mathcal{Z}}_a= -2H\mathcal{Z}_a -{1\over2}\,\rho^{(d)}\Delta^{(d)}_a+ {3\over2}\,a\rho^{(d)}\left(\dot{v}^{(d)}_a+Hv^{(d)}_a\right)- a{\rm D}_a\dot{\vartheta}^{(d)}- 2aH\,{\rm D}_a\vartheta^{(d)}\,,  \label{EdSdotcZ}
\end{equation}
respectively. The above monitors the linear evolution of linear peculiar velocities in a perturbed Einstein-de Sitter universe. Before proceeding to the  solution of the system, we remind the reader that the last term in Eq.~(\ref{EdSdotv}) and the last three terms on the right-hand side of (\ref{EdSdotcZ}) carry the input of the peculiar flux to the relativistic gravitational field.\footnote{There are more than one ways of deriving Eqs.~(\ref{EdSdotv}) and (\ref{EdSdotcZ}). Here, they have been obtained from the generalsed two-fluid formulae. Alternatively, one can confine to dust and start from the relativistic conservations laws of energy and momentum, or go to the equations describing structure formation (see~\cite{T,PT}, as well as \S~6.3 in~\cite{TPA} for details). In all cases, the key is accounting for the gravitational input of the peculiar flux.}

Differentiating (\ref{EdSdotv}) with respect to time, recalling that $3H^2=\rho^{(d)}$ and $\dot{H}=-3H^2/2$ in the Einstein-de Sitter background, using expression (\ref{EdSdotcZ}) and employing the linear commutation law $({\rm D}_af)^{\cdot}={\rm D}_a\dot{f}+H{\rm D}_af$ for first-order scalars, a straightforward calculation leads to the non-homogeneous differential equation
\begin{equation}
\ddot{v}^{d}_a+ H\dot{v}^{(d)}_a- {3\over2}\,H^2v^{(d)}_a= {1\over3aH}\left(\ddot{\Delta}^{(d)}_a+ 2H\dot{\Delta}^{(d)}_a- {3\over2}\,H^2\Delta^{(d)}_a\right)\,,  \label{EdSddotv}
\end{equation}
in complete agreement with~\cite{T,PT}. Isolating and solving the homogeneous left-hand side of the above leads to the power-law solution\footnote{The terms homogeneous/non-homogeneous refer to the nature of the differential equation and not to the homogeneity/inhomogeneity of the host space, which is both inhomogeneous and anisotropic at the linear level.}
\begin{equation}
v^{(d)}= \mathcal{C}_1t+ \mathcal{C}_2t^{-2/3}\,,  \label{hmv}
\end{equation}
given that $H=2/3t$ throughout the Einstein-de Sitter era~\cite{T}. Therefore, during this epoch, linear peculiar velocities grow as $v^{(d)}\propto t$. Moreover, this is the minimum growth-rate of the $v$-field, as ensured by a well-known theorem on differential equations. The latter states that the general solution of a non-homogeneous differential equation consists of the general solution of the corresponding homogeneous equation (of (\ref{hmv}) in our case) plus a particular solution of the full equation. Consequently, solving Eq.~(\ref{EdSddotv}) in full will make physical difference only if the extra mode grows faster than the fastest-growing mode of the homogeneous solution (\ref{hmv}).

One can also demonstrate the validity of the aforementioned theorem in practice. Solving only the homogeneous left-hand side of (\ref{EdSddotv}) amounts to neglecting the effects of the density gradients. Indeed, assuming that the density contrast evolves as $\Delta=\mathcal{C}_1t^{2/3}+\mathcal{C}_2t^{-1}$, namely as in the absence of peculiar motions, the right-hand side of (\ref{hmv}) vanishes identically and $v \propto t$. It is straightforward to show that if the density perturbations were to grow faster than $\Delta\propto t^{2/3}$, the peculiar-velocity field would also grow faster than $v\propto t$. For example, setting $\Delta\propto t$ on the right-hand side of (\ref{EdSddotv}), one obtains
\begin{equation}
v^{(d)}= \mathcal{C}_1t+ \mathcal{C}_2t^{-2/3}+ \mathcal{C}_3 t^{4/3}\,,  \label{fv1}
\end{equation}
where the extra mode has increased the velocity growth to $v^{(d)}\propto t^{4/3}$. On the other hand, if the density contrast grows slower than $\Delta\propto t^{2/3}$ (as $\Delta\propto t^{1/3}$ for instance), Eq.~(\ref{EdSddotv}) yields
\begin{equation}
v^{(d)}= \mathcal{C}_1t+ \mathcal{C}_2t^{-2/3}+ \mathcal{C}_3t^{2/3}\,.  \label{fv2}
\end{equation}
In this case, the additional mode makes no practical difference and the velocity field still grows as $v^{(d)}\propto t$. The latter is therefore the minimum growth-rate of linear peculiar velocities in the Einstein–de Sitter era, in agreement with the aforementioned theorem on differential equations.

\subsection{The late $\Lambda$-dominated epoch}\label{ssLL-DE}
According to the current cosmological paradigm, the $\Lambda$CDM model, the Einstein-de Sitter epoch is followed by a period of accelerated expansion, potentially driven by the gradual dominance of an underlying cosmological constant. Such a dramatic transition from universal deceleration to cosmic acceleration is bound to affect the evolution of peculiar motions.

Let us therefore consider an FRW background with positive cosmological constant and pressure-free matter (baryonic or/and CDM). Assuming that the latter has peculiar-velocity $v_a^{(d)}$ relative to the CMB frame, we perturb the aforementioned background. To linear order, the new environment has no effect on differential equation (\ref{EdSdotv}), but changes Eq.~(\ref{EdSdotcZ}) into
\begin{equation}
\dot{\mathcal{Z}}_a= -2H\mathcal{Z}_a -{1\over2}\,\rho^{(d)}\Delta_a^{(d)}+ a\left(3H^2+{1\over2}\,\rho^{(d)}-\Lambda\right) \left(\dot{v}_a^{(d)}+Hv_a^{(d)}\right)- a{\rm D}_a\dot{\vartheta}^{(d)}- 2aH{\rm D}_a\vartheta^{(d)}\,.  \label{dotcZ4}
\end{equation}
As in the Einstein-de Sitter case (see \S~\ref{ssEdSE} previously), the last three terms of the above carry the gravitational input of the peculiar flux.

The presence of a cosmological constant also modifies the background Friedmann equations, which now read
\begin{equation}
3H^2= \rho^{(d)}+ \Lambda \hspace{15mm} {\rm and} \hspace{15mm} \dot{H}= -{1\over2}\left(3H^2-\Lambda\right)\,.  \label{LambdaFRW1}
\end{equation}
Also, recalling that $\Omega_{\Lambda}=\Lambda/3H^2$, the above recast into the relations
\begin{equation}
\rho^{(d)}= 3H^2\left(1-\Omega_{\Lambda}\right) \hspace{15mm} {\rm and} \hspace{15mm} \dot{H}= -{3\over2}\,H^2\left(1-\Omega_{\Lambda}\right)\,,  \label{LambdaFRW2}
\end{equation}
respectively. Note that $0<\Omega_{\Lambda}<1$, with $\Omega_{\Lambda}\simeq0$ when matter dominates the energy density of the universe, and with $\Omega_{\Lambda}\simeq1$ when the total energy is dominated by the cosmological constant.

Differentiating (\ref{EdSdotv}) in terms of time, using all of the above and proceeding as in the Einstein-de Sitter case discussed previously, we arrive at
\begin{eqnarray}
\ddot{v}^{(d)}_a+ H\left(1+3\Omega_{\Lambda}\right)\dot{v}^{(d)}_a- {3\over2}\,H^2\left(1-3\Omega_{\Lambda}\right)v^{(d)}_a&=& {1\over3aH}\left[\ddot{\Delta}^{(d)}_a+ 2H\dot{\Delta}^{(d)}_a \right. \nonumber\\&&\left. -{3\over2}\,H^2\left(1-\Omega_{\Lambda}\right)\Delta^{(d)}_a\right]\,.  \label{Lddotv}
\end{eqnarray}
This non-homogeneous differential equation monitors the linear evolution of peculiar velocities in the pressureless matter of a perturbed FRW universe with nonzero cosmological constant. Clearly, when the latter vanishes, $\Omega_{\Lambda}=0$ and expression (\ref{Lddotv}) coincides with its Einstein-de Sitter counterpart (see Eq.~(\ref{EdSddotv}) in \S~\ref{ssEdSE}). In analogy with the preceding Einstein-de Sitter epoch, we will initially ignore the effect of the density gradients and solve analytically only the homogeneous left-hand side of (\ref{Lddotv}) for characteristic values of $\Omega_{\Lambda}$.\footnote{Linear density perturbations in the presence of $\Lambda$, but in the absence of peculiar velocities were studied in~\cite{VL}. If we ignore the peculiar-velocity perturbations as well, Eq.~(\ref{Lddotv}) here will agree with Eq.~(44) in~\cite{VL}.}

To begin with, suppose that the cosmological constant has an appreciable contribution to the total energy density of the universe, though without yet dominating over the matter component. In particular, let us set $\Omega_{\Lambda}=1/3$. Then, $H=1/t$ and the left-hand side of (\ref{Lddotv}) reads
\begin{equation}
t\ddot{v}^{(d)}_a+ 2\dot{v}^{(d)}_a= 0\,,  \label{Lddotvl}
\end{equation}
which solves to give
\begin{equation}
v^{(d)}= \mathcal{C}_1+ \mathcal{C}_2t^{-1}\,.  \label{Lv1}
\end{equation}
Accordingly, at the $\Omega_{\Lambda}=1/3$ threshold, linear peculiar-velocity perturbations no longer grow, but remain constant. This means that the peculiar-velocity growth has been suppressed even before the transition from decelerated expansion to cosmic acceleration.

Suppose now that $\Lambda$ has become the major contributor to the total energy density of the universe, but without fully dominating the universal dynamics. In particular, let us assume that $\Omega_{\Lambda}=2/3$, in which case the background Raychaudhuri equation (\ref{LambdaFRW2}b) integrates to give $H=2/t$. Then, the left-hand side of Eq.~(\ref{Lddotv}) becomes
\begin{equation}
t^2\ddot{v}^{(d)}_a+ 6t\dot{v}^{(d)}_a+ 6v^{(d)}_a= 0\,,  \label{Lddotv2}
\end{equation}
with
\begin{equation}
v^{(d)}= \mathcal{C}_1t^{-2}+ \mathcal{C}_2t^{-3}\,.  \label{Lv2}
\end{equation}
Therefore, when $\Omega_{\Lambda}=2/3$, linear peculiar-velocity perturbations decay as $v^{(d)}\propto t^{-2}$.

Finally, for completeness, let us also consider the case where the cosmological constant fully dominates the dynamics of the universe and set $\Omega_{\Lambda}=1$. In such an environment, the homogeneous left-hand side of differential equation (\ref{Lddotv}) takes the form
\begin{equation}
\ddot{v}^{(d)}_a+ 4H\dot{v}^{(d)}_a+ 3H^2v^{(d)}_a= 0\,,  \label{LddotmDel2}
\end{equation}
with $H\simeq H_0=$~constant throughout this phase of de Sitter-like  expansion (see Eq.~(\ref{LambdaFRW2}b) previously). The above accepts the solution
\begin{equation}
v^{(d)}= \mathcal{C}_1{\rm e}^{-H_0t}+ \mathcal{C}_2{\rm e}^{-3H_0t}\,,  \label{Lv3}
\end{equation}
showing exponential decay for the peculiar-velocity field during such a $\Lambda$-dominated epoch, namely at the $\Omega_{\Lambda}\rightarrow1$ limit.

Before closing this section, it is worth noting that, strictly speaking - and in accordance with the theory of differential equations (see \S~\ref{ssEdSE} before) - neglecting the effects of the density gradients and solving only the homogeneous part of Eq.~(\ref{Lddotv}) implies that solutions (\ref{Lv1}), (\ref{Lv2}) and (\ref{Lv3}) provide the maximum velocity suppression/decay. Nevertheless, even when the density gradients are included, the evolution of the $v$-field still exhibits an initial suppression of growth followed by a transition to decay. This happens even in the extreme (and physically unrealistic) scenario where $\Delta$ continues to evolve as in the preceding Einstein–de Sitter epoch, despite the increasing $\Omega_{\Lambda}$-contribution. Indeed, setting $\Delta\propto t^{2/3}$ on the right-hand side of (\ref{Lddotv}), one finds $v^{(d)}\propto t^{2/3}$ when $\Omega_{\Lambda}=1/3$, $v^{(d)}\propto t^{-1/3}$ when $\Omega_{\Lambda}=2/3$ and $v^{(d)}\propto {\rm e}^{-H_0t}$ when $\Omega_{\Lambda}=1$.

\begin{figure}[!tbp]\vspace{-7.5mm}
  \begin{subfigure}[b]{0.475\textwidth}
    \includegraphics[width=\textwidth]{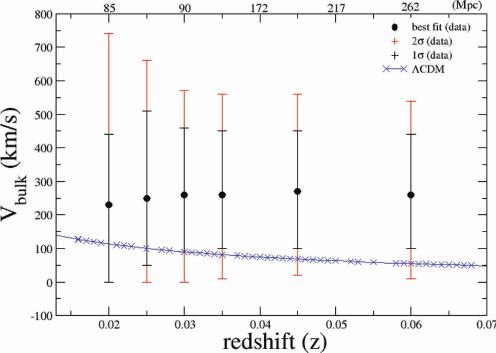}
    \caption{from Colin, et al~\cite{Cetal}}
    \label{fig:f1}
  \end{subfigure}
  \hfill
  \begin{subfigure}[b]{0.475\textwidth}
    \includegraphics[width=\textwidth]{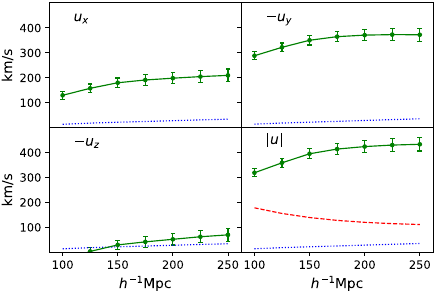}
    \caption{from Watkins, et al~\cite{Wetal}}
    \label{fig:f2}
  \end{subfigure}
  \caption{(a) The redshift profile of the bulk flow from~\cite{Cetal}. The peculiar velocity (black dots) systematically exceeds the blue line of the $\Lambda$CDM expectations, but its value also shows signs of decrease at lower redshifts. (b) The scale dependence of the bulk-flow velocity along the three axes and of its mean value (green lines), based on \textit{CosmicFlows4}~\cite{Wetal}. The red dashed line is the $\Lambda$CDM expectation. Note the profound disagreement between the predicted and the measured bulk velocities, as well as the clear decline in the magnitude of the latter at lower redshifts.}  \label{fig:CetalWetal}
\end{figure}

Overall, general relativity predicts a stronger linear growth of peculiar-velocity perturbations, than the Newtonian/quasi-Newtonian treatments, during the Einstein–de Sitter epoch. Assuming a subsequent phase of a gradually increasing $\Omega_{\Lambda}$-contribution, this growth first slows down and eventually transitions into decay. Nevertheless, part of the earlier velocity growth is expected to persist to the present epoch. Consequently, one would anticipate higher bulk velocities at earlier times - exceeding those predicted by Newtonian-based $\Lambda$CDM models - followed by a gradual decline toward lower redshifts. In qualitative terms, the resulting peculiar-velocity profile would resemble those reported in~\cite{Cetal,Wetal} (see also Fig.~\ref{fig:CetalWetal} here).

\section{Discussion}\label{sD}
The fast and deep bulk peculiar flows, with sizes up to several Mpc and velocities up to several km/sec, that have been repeatedly reported in the literature remain an open puzzle. While the surveys seem to agree with the direction of the reported flows, they differ significantly in their inferred speeds and sizes. On scales up to $\sim100$~Mpc, the observations are broadly consistent with the $\Lambda$CDM prediction of velocities of up to $\sim100$~km/s (e.g.~see~\cite{ND}–\cite{Fetal}). However, on larger scales, they increasingly deviate from these expectations (e.g.~see~\cite{WFH}–\cite{WHD}). To this category, one may also add the so-called ``dark flows'' (e.g.~see~\cite{Ketal}–\cite{AF}), whose amplitudes and scales significantly exceed the $\Lambda$CDM limits (see~\cite{TPA} for a recent review and additional references).

As long as it remains unresolved, the bulk-flow puzzle has the potential to challenge the foundations of the current cosmological model. Having said that, the $\Lambda$CDM predictions are based entirely on Newtonian theory, which argues for a relatively weak linear growth of peculiar velocities ($v\propto t^{1/3}$) during the Einstein–de Sitter era. Such a growth-rate is insufficient to account for the observed fast and deep bulk flows, particularly when a subsequent phase of accelerated expansion is taken into consideration~\cite{Detal}.

In a series of papers culminating in~\cite{T,PT} (see also \cite{TPA}), a fully relativistic treatment of cosmological peculiar velocities was developed. These studies revealed that general relativity predicts a stronger linear growth for peculiar velocities ($v\propto t$), leading naturally to faster and deeper bulk flows today. This difference arises because a consistent relativistic treatment of peculiar motions must also include their flux contribution to the gravitational field. Recall that peculiar flows imply moving matter and therefore nonzero energy flux. In general relativity, energy fluxes gravitate, as they also contribute to the energy-momentum tensor. Purely Newtonian analyses neglect this effect by default, while the quasi-Newtonian approaches suppress it due to the strict constraints they impose on the perturbed spacetime. As a result, the quasi-Newtonian treatments lose essential relativistic features and reduce to Newtonian conclusions. Although already noted in \S~6.8.2 of~\cite{EMM}, the quasi-Newtonian limitations were not fully appreciated until the proper relativistic framework was developed (see \cite{TPA} for a detailed discussion).

In the present work, we have extended the recent relativistic analysis of~\cite{T,PT} to a late era of progressive $\Lambda$-domination, in order to assess its impact on the evolution of peculiar-velocity perturbations. Previous studies, combining Newtonian theory with numerical methods, also concluded that peculiar velocities grow slower during the $\Lambda$-dominated phase than in the preceding Einstein–de Sitter epoch~\cite{Detal}. By and large, this behavior is to be expected, as cosmic acceleration generally suppresses the growth of perturbations. For example, the analysis of peculiar velocities during a de Sitter inflationary phase predicted an exponential decay for the $v$-field~\cite{MaT}. Here, by tracking the evolution of linear peculiar velocities in a universe with a gradually increasing $\Omega_\Lambda$-contribution, we found that the latter initially suppresses the Einstein–de Sitter growth and eventually reverses it into decay. Qualitatively, this suggests that bulk velocities should exceed the (Newtonian-based) $\Lambda$CDM expectations at intermediate stages, before declining at lower redshifts. Interestingly, such velocity profiles have been reported in~\cite{Cetal,Wetal}. All this raises the possibility that the observed fast and deep bulk flows may not signal a breakdown of the $\Lambda$CDM model, but rather reflect the limitations of applying an inadequate (Newtonian) gravitational framework to the study of cosmological peculiar motions.

Finally, our analysis also provides a useful framework for future numerical studies of cosmological peculiar velocities. Such investigations should make it possible to track the evolution of the velocity field across the transition from the Einstein–de Sitter epoch to the more recent phase of gradual $\Omega_{\Lambda}$-domination, namely during stages where analytical solutions are not available. Since typical numerical studies are performed in a specific gauge, care must be taken to avoid gauge choices that may inadvertently eliminate the genuinely relativistic contribution of the peculiar flux to the gravitational field. In this context, the gauge invariance of our analysis (as well as that of the earlier relativistic studies of~\cite{TT}–\cite{PT}) is a significant advantage, ensuring that the associated solutions are free from spurious modes and gauge-related ambiguities.\\

\textbf{Acknowledgements:} This work was supported by the Hellenic Foundation for Research and Innovation (H.F.R.I.), under the ``First Call for H.F.R.I. Research Projects to support Faculty members and Researchers and the procurement of high-cost research equipment Grant'' (Project Number: 789).

\end{document}